# Ultralow-noise microwave oscillator via optical frequency division with a co-self-injection-locked miniature Fabry–Pérot reference

Runlin Miao[1]†*, Chao Zhou[2]†, Pan Han[3]†, Mingxin Yang[3], Xing Zou[3], Ke Wei[4,5]*, Ke Yin[3], Tian Jiang[4,5]*

[1]National Innovation Institute of Defense Technology, Academy of Military Sciences PLA China, Beijing, China
[2]College of Computer Science and Technology, National University of Defense Technology, Changsha, China
[3]College of Advanced Interdisciplinary Studies, National University of Defense Technology, Changsha, China
[4]College of Science, National University of Defense Technology, Changsha, China
[5]Hunan Research Center of the Basic Discipline for Physical States, National University of Defense Technology, Changsha, China
†These authors contributed equally to this work.
*Corresponding author. Email: mrl1123@126.com, weikeaep@163.com, tjiang@nudt.edu.cn

## Abstract

Optical frequency division (OFD) provides the purest microwaves by down-converting the stability of optical cavity references. State-of-the-art references typically rely on electronic co-Pound-Drever-Hall locking to ultrahigh-Q microresonators—a complex approach that introduces servo bumps and increases footprint. Alternatively, optical co-self-injection-locking (co-SIL) offers inherent simplicity but is limited by the large thermo-refractive noise and confined mode volumes of integrated cavities. Here, we demonstrate a two-point OFD-based microwave oscillator that combines an ultrahigh-Q miniature Fabry–Pérot cavity with optical co-SIL. Leveraging its low relative phase noise optical reference and combing with an integrated soliton microcomb, the system generates a microwave with phase noise of −147 dBc/Hz at 4 kHz offset (scaled to 10 GHz)—performance rivalling most electronically stabilized systems. This work marries the superior noise floor of ultrahigh-Q cavities with the simplicity of optical locking, providing a compact, cost-effective, and field-deployable path to pure microwaves for next-generation communications, radar and metrology.

## Introduction

Generation of spectrally pure microwave and millimeter-wave signals is foundational to modern technologies such as 5G/6G wireless communications (*1*), radar (*2*), radio astronomy (*3*) and sensing (*4*). The performance requirements in these fields, especially close to the carrier, often exceed the capabilities of conventional electronic oscillators, which suffer from increased noise at high frequencies and limited quality factors (*5*). Photonic synthesis, like optical frequency comb (*6, 7*), dual-frequency lasers (*8*), optoelectronic oscillators (*9-11*) and Brillouin oscillators (*12, 13*), overcomes these challenges by leveraging optical cavities with extremely low loss and high-quality factors (*14-16*). Among these, optical frequency division (OFD) stands out, transferring the exceptional fractional stability of an optical reference to the microwave domain via an optical frequency comb. Conventional OFD employs bulky, high-finesse optical cavities coupled to octave-spanning mode-locked laser combs, delivering remarkable performance but compromising portability and cost, thus restricting use to laboratory settings (*17, 18*). For practical field deployment with strict size, weight, power and cost (SWaP-C) constraints, the integrated two-point OFD (2P-OFD) architecture has become pivotal (*19-23*). This approach avoids the need for octave-spanning combs by phase-locking a chip-based frequency comb to two laser references derived from a single compact cavity.

The ultimate phase noise performance of a 2P-OFD system is dictated by its dual-laser optical references. State-of-the-art implementations predominantly use the co-Pound–Drever–Hall (co-PDH) technique to electronically lock external-cavity lasers to ultrahigh-Q microresonators (*24, 25*). Although achieving remarkable stability, these systems necessitate complex feedback loops involving narrow-linewidth lasers (linewidth <1 kHz), electro-optic modulators, and high-speed servo electronics, which increase cost, footprint, and introduce characteristic servo bumps at high offset frequencies (~ 200 kHz). In parallel, all-optical stabilization methods have emerged as a fundamentally simpler alternative, eliminating active electronic feedback and laser modulation while enabling full-offset-frequency noise suppression. These include mechanisms such as optical parametric oscillation (OPO) (*26*) and co-self-injection locking (co-SIL) (*27*), typically implemented on photonic-integrated platforms such as silicon nitride ($Si_3N_4$) ring or spiral resonators. Since the dual-laser noise is negatively correlated with the quality factor (Q) and the mode volume, limited Q (typically <$10^8$) and small mode volumes (< $2.5\times10^{-11}$ $m^3$) in these on-chip cavities restrict reported optically locked references to phase noise between −57 and −94 dBc/Hz @ 10 kHz offset, with significant degradation at lower offsets due to the ambient temperature perturbations and random vibrations. Beyond integrated resonators, ultrahigh-Q platforms such as optical fibers (*28*), whispering gallery mode resonators (*29*), and miniature Fabry-Pérot (mini-FP) cavities (*30*) also support optical locking. Mini-FP cavities are particularly notable, offering large mode volume (>$10^{-10}$ $m^3$), ultrahigh Q (>$10^9$), low thermal absorption, low thermal refractive noise, and straightforward coupling (*31*). Recent advances in co-SIL to mini-FP cavities have demonstrated ultralow relative phase noise lasers (*30, 32*). Nevertheless, the crucial integration of such a superior optical reference with a chip-based frequency comb to perform 2P-OFD, thereby transforming exceptional optical stability into a spectrally pure microwave signal, remains an open challenge and a compelling opportunity.

In this work, we bridge this gap by demonstrating the first photonic microwave oscillator based on 2P-OFD using an optically stabilized air-gap mini-FP cavity. We generate an optical reference with ultralow phase noise (−106 dBc/Hz @ 10 kHz offset) by co-SIL of two commercial on-chip distributed feedback (DFB) lasers (linewidth > 100 kHz) to a single mini-FP cavity (mode volume ~ $3.5\times10^{-10}$ $m^3$), achieving full offset frequency common-mode noise suppression. This reference is then integrated with an on-chip microcomb in a 2P-OFD scheme to generate ultralow-noise U-band microwave signals at ~50 GHz. The resulting oscillator significantly outperforms state-of-the-art optically locked references (*26, 27*), achieving a record phase noise of −99 dBc/Hz @ 100 Hz offset and −147 dBc/Hz @ 4 kHz offset (scaled to 10 GHz)—performance comparable to, or better than, most co-PDH-based systems (*31, 33-36*). Our work successfully marries the outstanding noise performance of discrete ultrahigh-Q cavities with the architectural simplicity of optical locking for 2P-OFD, without requiring bulky and expensive narrow-linewidth fiber lasers, frequency locking components, or complex control modules. This makes the system readily manufacturable and well suited for large-scale, real-world deployment.

## Results

### Concept of the 2P-OFD scheme

Figure 1 illustrates the architecture and operating principle of our photonic microwave oscillator, which synthesizes an ultralow-noise 50 GHz signal through a 2P-OFD scheme anchored by a co-SIL mini-FP cavity. The system comprises two core subsystems: an all-optically stabilized dual-laser reference (Fig. 1A) and a 2P-OFD phase-locking module that bridges the optical and microwave domains (Fig. 1B). The foundational stability originates from the co-SIL reference, where two commercial DFB lasers (optical frequencies $f_A$ ~ 192.3 THz and $f_B$ ~ 194.8 THz) are simultaneously locked via optical feedback to distinct longitudinal modes of a single air-gap mini-FP cavity (*37*). This passive, all-optical mechanism suppresses common-mode noise between the lasers, yielding a pristine ~2.5 THz optical beatnote without the complexity, cost, and servo bumps associated with electronic feedback techniques such as co-PDH locking. This exceptional optical stability is then transferred to the microwave domain via the 2P-OFD scheme. A soliton microcomb, with a repetition rate $f_r \approx 50$ GHz, is generated by pumping a $Si_3N_4$ microresonator with a third, low-cost DFB laser (*38*). In the 2P-OFD loop, specific comb teeth ($f_m$, $f_n$) are heterodyned with the two co-SIL reference lasers, producing two radio-frequency beat notes $\Delta_1$ and $\Delta_2$. These are mixed to generate an intermediate-frequency error signal $f_{IF} = |\Delta_1 − \Delta_2|$, which inherently cancels the comb's carrier-envelope offset frequency. By phase-locking $f_{IF}$ to a stable local oscillator ($f_{LO}$) and feeding the error signal back to the pump laser current, the microcomb's repetition rate $f_r$ is rigidly tied to the optical reference. This works because varying the laser current changes the pump frequency, and thus $f_r$ is tuned via Raman induced soliton self-frequency shift and dispersive-wave recoil (*39*). At this point, photodetection of the locked microcomb yields a low-noise microwave signal, with repetition rate of $f_r = \frac{f_B - f_A - f_{LO}}{n-m}$.

The theoretical phase-noise suppression of this architecture, visualized in Fig. 1C, proceeds via three cascaded stages. First, SIL to the ultrahigh-Q mini-FP cavity suppresses the inherent frequency noise of each free-running DFB laser. The noise-reduction factor (NRF) scales as NRF $\propto$ $(Q/Q_{LD})^2$, where Q (~$3.5\times10^9$, as characterized below) is the cavity quality factor and $Q_{LD}$ (~$10^4$) is that of the free-running laser. This provides ~60 dB of noise suppression across the offset frequency band. Second, because both lasers are locked to the same physical cavity, their fluctuations are strongly correlated. The relative phase noise of their ~2.5 THz beatnote thus benefits from common-mode rejection, yielding an additional 20–40 dB of suppression. Third, the 2P-OFD down-converts this stable optical reference to the microwave domain. With the $f_{LO}$ noise negligible compared to the optical beat $f_B - f_A$, the single-sideband phase noise of the generated microwave is reduced by a factor of $N^2$, where $N = (f_B - f_A)/f_r \approx 50$ is the division ratio. This corresponds to $10\log_{10}(N^2) \approx 34$ dB of suppression within the loop bandwidth. In total, the architecture theoretically enables over 114 dB of cumulative noise suppression, translating the fundamental noise floor of the ultrahigh-Q mini-FP cavity directly into the microwave regime.

**Optical reference via co-SIL to a mini-FP cavity**

The phase noise foundation of our 2P-OFD system is established by a compact, co-SIL dual-laser reference module. As schematically illustrated in Fig. 2A, the module integrates two on-chip DFB lasers, an air-gap mini-FP cavity, and other micro-optical components into a fully electric-driven package measuring only $75 \times 40 \times 28$ mm$^3$ (see Methods for details), highlighting its potential for field deployment. Specifically, the outputs of the two DFB lasers are first combined and then split into two primary paths. The first path is directed towards the mini-FP cavity to provide the optical feedback necessary for SIL. The second path is routed through an optical isolator, delivering a combined output of several milliwatts for subsequent use in the 2P-OFD system (see optical spectra in Fig. 2C). For the locking feedback path, we implement a compact SIL configuration by slightly tilting the mini-FP cavity to utilize reflected high-order transverse modes, rather than the commonly used transmitted fundamental modes (*37*). This approach effectively filters out non-resonant reflections via free-space filtering while enabling sufficient backscattering for stable locking. The light transmitted through the mini-FP cavity is critically utilized for locking control and verification. It is subsequently split into two beams: one is then split by a dichroic mirror and detected by photodetectors (PDs) to monitor the transmission dip, providing the real-time signal for initiating and maintaining the SIL state; the other is coupled out as free-space light, enabling direct observation and characterization (e.g., beam profiling) of the specific high-order transverse mode to which each laser is locked. By tuning the silicon phase shifter to adjust the optical feedback phase, simultaneous SIL of both DFB lasers to selected high-order cavity modes is achieved, as evidenced by a pronounced increase in transmitted light intensity (Fig. 2D). The measured locking range exceeds 4 GHz (estimated from the PD signal width), facilitating robust and persistent co-SIL that is maintained for several hours in an open laboratory environment without active stabilization.

The exceptional performance of this reference stems from the ultrahigh quality factor of the mini-FP cavity. Ring-down measurement (see Supplementary Note 1 for setup) yields an amplitude decay lifetime of $\tau \sim 5.75$ μs (Fig. 2B), corresponding to a Q-factor of $Q = \pi c \tau / \lambda = 3.5 \times 10^9$, with $c$ the speed of light and $\lambda$ the laser wavelength. Such ultrahigh $Q$ dramatically enhances the noise reduction factor of SIL. The laser frequency noise and linewidth, characterized via a delayed self-homodyne method (Supplementary Note 2), show remarkable improvement upon locking. As shown in Fig. 2E, both SIL states suppress the frequency noise by over 60 dB across the measured offset range compared to the free-running state. The Lorentzian linewidth narrows from 16.0 kHz to 90.7 mHz, and the integral linewidth reduces from 188.4 kHz to 144.5 Hz, significantly surpassing the performance of a commercial Koheras Basik E15 fiber laser. The relative phase noise between the two co-SIL lasers, measured by combining a Menlo-comb-assisted technique (for offsets <1 kHz) and two-wavelength delayed interferometry (for offsets >1 kHz) (Methods and Supplementary Note 3), reaches a record-low value of -106 dBc/Hz at a 10 kHz offset for the 2.5 THz beat note (Fig. 2F, yellow curve). For comparison, the phase noise of a single SIL laser is also shown (blue curve), demonstrating that co-SIL provides an additional 20~40 dB suppression from 10 Hz to 10 kHz offsets through common-mode noise rejection. Critically, no servo bump is observed, benefiting from the all-optical locking mechanism.

Fig. 2G places the performance of our co-SIL dual-laser reference in context with state-of-the-art systems. It achieves the lowest relative phase noise among all optically locked references across a broad offset range from 10 Hz to 1 MHz. Notably, at low offsets (10 Hz-1 kHz), its performance matches that of the electronically stabilized (co-PDH) references, owing to the large mode volume and vibration insensitivity of our mini-FP cavity. At higher offsets, it avoids the characteristic servo bumps introduced by electronic feedback loops. The ultimate noise floor is likely limited by the cavity's fundamental noise, arising from the thermo-dynamics mediated refractive index fluctuations of the host material (*40*). This co-SIL reference module thus successfully combines ultralow noise performance with architectural simplicity and compactness, forming the ideal cornerstone for the subsequent 2P-OFD system.

**Ultralow-noise 50 GHz microwave generation via 2P-OFD**

Having established the ultralow-noise co-SIL dual-laser reference, we now integrate it with a chip-based soliton microcomb to perform the 2P-OFD and generate a spectrally pure microwave signal. The complete experimental setup is detailed in Supplementary Note 5. A key aspect of our approach is the soliton microcomb generation scheme, which employs a dual-mode $Si_3N_4$ microresonator (Supplementary Note 4). Pumping two adjacent modes enables self-thermal compensation, significantly broadening the soliton existence regime (*41, 42*) and allowing deterministic single-soliton generation using only a low-cost DFB pump laser with slow current tuning, thereby eliminating the need for expensive narrow-linewidth lasers with fast piezoelectric actuators, simultaneously cutting link cost and reducing form factor. The generated single-soliton microcomb exhibits a repetition rate of ~50 GHz and a 3 dB optical bandwidth of approximately 36.3 nm (Fig. 3A). This bandwidth fully covers the 20 nm (2.5 THz) separation of the co-SIL reference lasers, ensuring sufficient comb teeth power for beating to generate the error signal. This error signal is then phase-locked to the LO, enabling frequency division of 2.5 THz to 50 GHz. A dramatic linewidth narrowing is observed upon locking: the free-running soliton beatnote shows a broad, noisy spectrum (Fig. 3B), whereas the locked 50.015 GHz signal displays a coherent, narrow-linewidth carrier with a pronounced servo bump at ~200 kHz offset, confirming the locking bandwidth (Fig. 3C).

The phase noise performance of the resulting 50 GHz microwave signal, measured using a phase noise analyzer, is presented in Fig. 3D. The phase noise of the locked microcomb closely follows the projected noise floor of the co-SIL reference, scaled down by the theoretical $10\log_{10}(50^2)$ ~34 dB division factor (purple dotted curve). This demonstrates successful noise transfer via optical frequency division. Compared to its free-running state, the locked signal achieves over 60 dB of noise suppression from 10 Hz to 4 kHz offset, reaching a minimum of -133 dBc/Hz at 4 kHz. For direct comparison with standard microwave oscillators, the noise is scaled to a 10 GHz carrier (green dotted curve, vertical scale offset by $10\log_{10}[50\ \text{GHz}/10\ \text{GHz}]^2 = 14$ dB), yielding outstanding performance of $-147$ dBc/Hz @ 4 kHz and $-144$ dBc/Hz @ 10 kHz. The phase noise above ~6 kHz offset is dominated by the in-loop error signal ($f_{IF}$) and is limited by the ~200 kHz servo bandwidth, a constraint primarily attributed to the relatively broad linewidth of the DFB pump laser, which degrades servo signal quality and thus enhances residual noise (see Supplementary note 2 for DFB linewidth result). By further improving the signal-to-noise ratio of the intermediate frequency signal and increasing the servo locking bandwidth, the servo bump can be shifted toward higher offset frequency, thereby reducing low-frequency phase noise.

To characterize the stability of the low-noise 50-GHz microwave signal, a waterfall plot measurement was performed using an electronic spectrum analyzer before and after locking (Fig. 3E). It is evident that the locked frequency exhibits extremely long-term high stability, in contrast to the ±20 kHz random jitter of the free-running signal. The fractional frequency stability, quantified by the Allan deviation measured with a phase noise analyzer referenced to a rubidium clock, is shown in Fig. 3F. The free-running signal exhibits an Allan deviation of $2.1\times10^{-8}$ at 1 s averaging time. After phase-locking, it reaches a minimum of $3\times10^{-12}$ at 63 ms. Beyond this averaging time, due to the long-term random environmental temperature drift affecting the mini-FP cavity's effective length, the Allan deviation gradually increases to $4.3\times10^{-11}$ at 1 s, which is also the stability of the 2.5 THz frequency reference generated by the co-SIL dual lasers.

**Noise dependence on co-SIL transverse modes**

The performance of the 2P-OFD system is intrinsically linked to the characteristics of the co-SIL reference, particularly the specific transverse modes to which the two DFB lasers are locked. Since the mini-FP cavity supports multiple transverse modes within a single free spectral range, we systematically investigate—for the first time, to our knowledge—how the choice of locking modes influences the relative phase noise of the dual lasers and thus the final microwave output. To this end, we compare three distinct co-SIL scenarios using two high-order transverse mode families, $TEM_{02}$ and $TEM_{13}$, as examples: (1) both lasers locked to the same $TEM_{02}$ mode; (2) both locked to the same $TEM_{13}$ mode; and (3) one laser locked to $TEM_{02}$ and the other to $TEM_{13}$. The targeted modes are selected by fine-tuning the laser currents and the phase through silicon phase shifters. The successful locking to the desired modes is confirmed by imaging the free-space output beam profiles, as shown in Fig. 4A.

The relative phase noise between the two co-SIL lasers under these three scenarios is measured and compared in Fig. 4B. A key finding is that locking both lasers to the same transverse mode family (either $TEM_{02}$ or $TEM_{13}$) yields identical, minimal relative phase noise, demonstrating optimal common-mode noise rejection. In contrast, when the lasers are locked to different mode families ($TEM_{02}$ & $TEM_{13}$), the common-mode noise suppression is significantly compromised, especially at higher offset frequencies. The relative phase noise in this case is 20 dB and 16 dB higher at 10 Hz and 10 kHz offsets, respectively, compared to the same-mode-family locking. This degradation is attributed to a reduced spatial overlap of the optical beams in different transverse modes, which diminishes the effectiveness of common-mode suppression within the shared cavity.

This dependence directly translates to the performance of the 2P-OFD-generated microwave. Figure 4C shows the single-sideband phase noise of the locked 50 GHz signal for the three co-SIL scenarios. The results fully track the relative phase noise of the co-SIL references (scaled by 34 dB). Consequently, the microwave phase noise is significantly lower and consistent when the dual lasers are locked to the same mode family, whereas locking to different modes leads to substantially degraded performance. This clearly establishes that locking to the same transverse mode family is critical for maximizing common-mode noise rejection and achieving the ultralow phase noise reported in Fig. 3D.

**Discussion**

Our work demonstrates a photonic microwave oscillator that successfully integrates an optically stabilized, discrete ultrahigh-Q mini-FP cavity with a chip-based soliton microcomb via a 2P-OFD architecture. This system delivers exceptional close-to-carrier phase noise, rivaling the performance of most electronically stabilized references, while simultaneously offering a dramatically simplified and more compact architecture. As detailed in Fig. 5 and Table 1, our oscillator delivers the best performance reported to date for optically referenced microwave sources at offset frequencies below 10 kHz. The phase noise of the co-SIL dual-laser reference itself reaches an ultralow -106 dBc/Hz at 10 kHz offset for a 2.5 THz beat note. When divided down, it enables a 50 GHz microwave with −99 dBc/Hz at 100 Hz offset and −147 dBc/Hz at 4 kHz (scaled to 10 GHz), surpassing all prior optically stabilized systems in this offset range and matching the performance of advanced co-PDH references near the carrier. This result definitively proves that the superior noise floor of a discrete ultrahigh-Q cavity can be fully transferred to the microwave domain using a purely optical locking technique, without the complexity of PDH electronics. A comprehensive evaluation, however, reveals a key trade-off inherent to our current implementation. The complete phase noise spectrum (see Supplementary Fig. S8) shows that the phase-locked loop used to lock the microcomb to the optical reference introduces a pronounced servo bump near 200 kHz offset. This degrades the high-offset-frequency (>10 kHz) noise performance compared to other optical references. This limitation stems primarily from the use of a low-cost DFB laser for microcomb pumping, whose relatively broad linewidth and high-frequency current tuning noise reduce the signal-to-noise ratio and bandwidth of the electronic feedback control.

The system's cost-effectiveness and practical scalability also constitute a major advantage. The co-SIL reference module eliminates the entire suite of specialized, high-value components required for co-PDH locking, including two ultra-narrow-linewidth external-cavity lasers, electro-optic and acousto-optic modulators, stable radio frequency local oscillators, and high-speed servo controllers. Instead, it employs only two standard telecommunication-grade DFB laser chips and a passively stable mini-FP cavity. Furthermore, for microcomb generation, we replace an expensive, kHz-linewidth fiber laser and its associated rapid piezoelectric scanning system (*33, 43-45*) or a full suite of auxiliary laser devices (*46, 47*) with a single low-cost DFB laser, leveraging a dual-mode thermal compensation scheme for soliton initiation. Consequently, the core photonic engine of our oscillator comprises merely three III-V DFB chips, one mini-FP cavity, and one $Si_3N_4$ microresonator—all components compatible with standard batch fabrication and assembly processes, paving the way for scalable manufacturing.

Future improvements can directly address the current high-frequency noise limitation without compromising these cost and integration benefits. Replacing the electronic feedback with an all-optical Kerr synchronization scheme could extend the locking bandwidth beyond 30 MHz (*34, 48-50*), fundamentally suppressing the servo bump and reach phase noise as low as −154 dBc/Hz @ 10 kHz @ 10 GHz. Alternatively, electrical feedforward techniques could actively cancel this noise (*51*). Furthermore, the phase noise suppression factor, $N^2$, can be directly increased by utilizing co-SIL lasers with a wider frequency separation within the microcomb's bandwidth, thereby dividing down a larger optical reference frequency for greater noise reduction. Further system simplification will be attainable by integrating the recently reported mini-FP cavity interface with photonic integrated circuits (*52, 53*). Additionally, on-chip optical amplifiers via erbium-doped $Si_3N_4$ (*54*) or parametric waveguides (*55,*

*56*) can replace the bulky erbium-doped fiber amplifier. The integration of these advancements promises the development of next-generation photonic microwave oscillators that combine ultralow close-in phase noise, clean high-offset spectra, and full integrability for field-deployable applications in communications, radar, and timing.

**Materials and Methods**

**Packaging of co-SIL dual lasers with mini-FP cavity**

The entire co-SIL dual-laser module is packaged in a box with dimensions of 75 × 40 × 28 mm$^3$. For the air-gap mini-FP cavity, it is mechanically constructed by mounting a commercial plane mirror (diameter: 6.35 mm, thickness: 1 mm) and a concave mirror (diameter: 6.35 mm, thickness: 2.3 mm, radius of curvature: 500 mm) on a hollow Zerodur spacer (~15 mm long, corresponding to a free spectral range (FSR) of ~9.85 GHz). Then, the two DFB laser diodes (in chip-on-carrier (COC) configuration) are mounted on a substrate. The mini-FP cavity and all other micro-optical components—including beam splitters, high-reflection mirrors, dichroic mirrors, monitor photodetector chips, silicon sheets, and optical isolators (ISO)—are also integrated onto the same substrate, enabling optical axis alignment and thus ensuring high stability and reliability. This module employs two thermoelectric cooler (TEC) temperature controllers: one for temperature regulation of one individual laser chip, and the other for global temperature control of the other laser chip and the entire substrate. The SIL feedback phase can be fine-tuned via resistive heating of the silicon sheet in the feedback loop. Notably, all components and assembly techniques in this module are consistent with those in the optical communication industry, enabling the mass production of the co-SIL dual-laser module when required.

**Characterization of co-SIL dual-laser references' relative phase noise**

To measure the relative phase noise of dual-laser references with a large frequency spacing (~2.5 THz), two setups were employed for stitching measurements across different offset frequencies (see Supplementary Note 3 for details). For offset frequencies ≤ 1 kHz, a Menlo comb-assisted scheme was adopted. Specifically, the carrier-envelope offset frequency ($f_{ceo}$) of the Menlo comb was first locked via self-referencing, then the Menlo comb was phase-locked to the 1560 nm laser—synchronizing its noise characteristics across all comb teeth. Finally, beating the mode-locked comb with the 1540 nm laser generated a signal below 100 MHz, whose phase noise enabled retrieval of the dual lasers' relative phase noise. For offset frequencies > 1 kHz, the widely used two-wavelength delayed interferometry (TWDI) technique was utilized. The dual lasers first underwent sub-coherence measurement via an unbalanced MZI. After wavelength division multiplexing (WDM) and PD detection, two separate 40 MHz electrical signals (containing individual laser phase jitter) were obtained. Subsequently, one signal was up-converted by a 70 MHz LO, followed by down-conversion with the other 40 MHz signal. The resulting 70 MHz signal thus contained the relative phase noise of the dual lasers. Combining these two methods enables accurate relative phase noise measurement of the dual lasers over a 10 Hz–1 MHz offset frequency range.

**Acknowledgments**

The $Si_3N_4$ chips used in this work were fabricated by Qaleido Photonics.

**Funding:**

National Natural Science Foundation of China (Grant No. 62522518)
The Science Fund for Distinguished Young Scholars of Hunan Province (Grant No. 2024JJ2054)
National Natural Science Foundation of China (Grant No. 12474351)

**Author contributions:**

Conceptualization: TJ, KW, RLM
Resources: RLM, CZ, PH
Formal analysis: RLM, XZ
Investigation: RLM, MXY
Funding acquisition: KW, TJ
Project administration: KY, KW, TJ
Writing – original draft: RLM, CZ, PH
Writing – review & editing: RLM, KY, KW, TJ

**Competing interests:**

The authors declare that they have no competing interests.

**Data and materials availability:**

The data that supports the plots within this paper and other findings of this study are available from the corresponding author upon reasonable request.

**Figures and Tables**

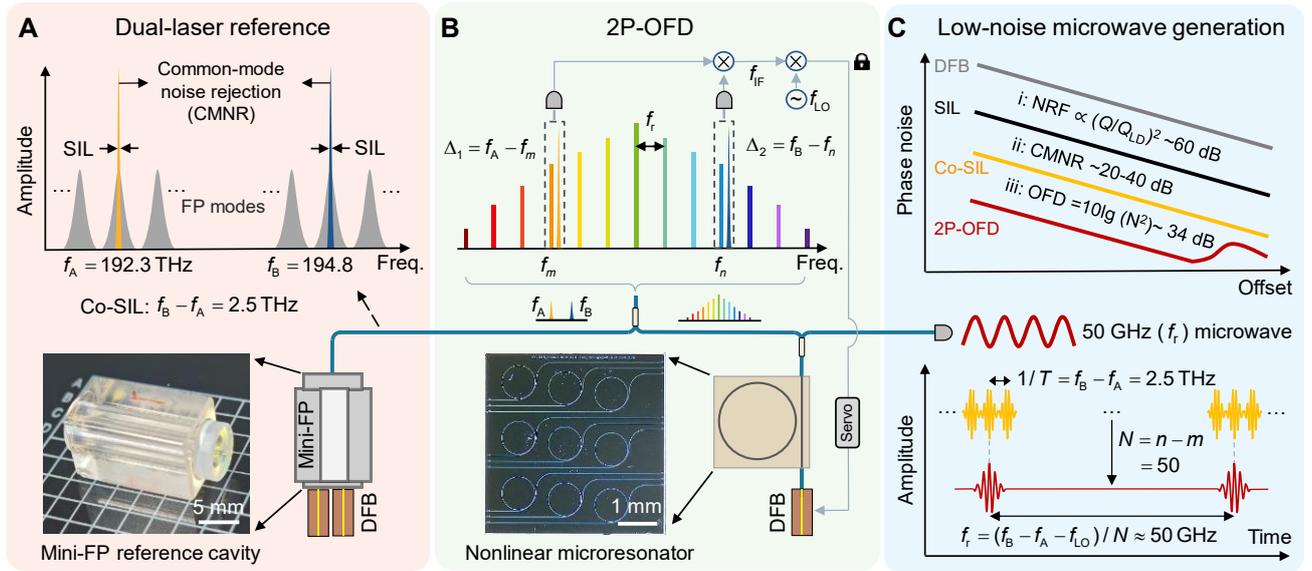

**Fig. 1. Operating principle of mini-FP based 2P-OFD for low-noise microwave generation. (A)** Ultra-stable optical reference via co-self-injection-lock (co-SIL). Two DFB lasers ($f_A$ and $f_B$) are optically locked to distinct longitudinal modes of a single air-gap mini-FP cavity through self-injection feedback, generating a stable ~2.5 THz beatnote with inherent common-mode noise suppression. A photograph of the mini-FP cavity is shown. **(B)** 2P-OFD phase-locking for noise transfer. A soliton microcomb (repetition rate $f_r \sim$ 50 GHz), generated by pumping a $Si_3N_4$ microresonator with a third DFB laser, is phase-locked to the co-SIL reference. The beatnotes ($\Delta_1$ and $\Delta_2$) between specific comb teeth ($f_m$ and $f_n$) and the two reference lasers are mixed to produce an intermediate-frequency error signal ($f_{IF}= |\Delta_1 - \Delta_2|$). This signal is compared with a local oscillator ($f_{LO}$) and fed back to the pump laser current, locking $f_r$ and thereby transferring the optical stability to the microwave domain. A photograph of the $Si_3N_4$ microresonator is shown. **(C)** Microwave generation and noise suppression. The locked microcomb is photodetected to produce a pure 50 GHz (U-band) microwave signal. Top: Conceptual illustration of the phase noise reduction achieved at each stage, highlighting the >114 dB suppression from the free-running state to the final divided microwave. Bottom: Diagram of the optical-to-microwave frequency division process via the 2P-OFD scheme, with a division factor of $N$ = 50, corresponding to a theoretical ~34 dB ($10\log_{10}(N^2)$) noise reduction within the locking bandwidth.

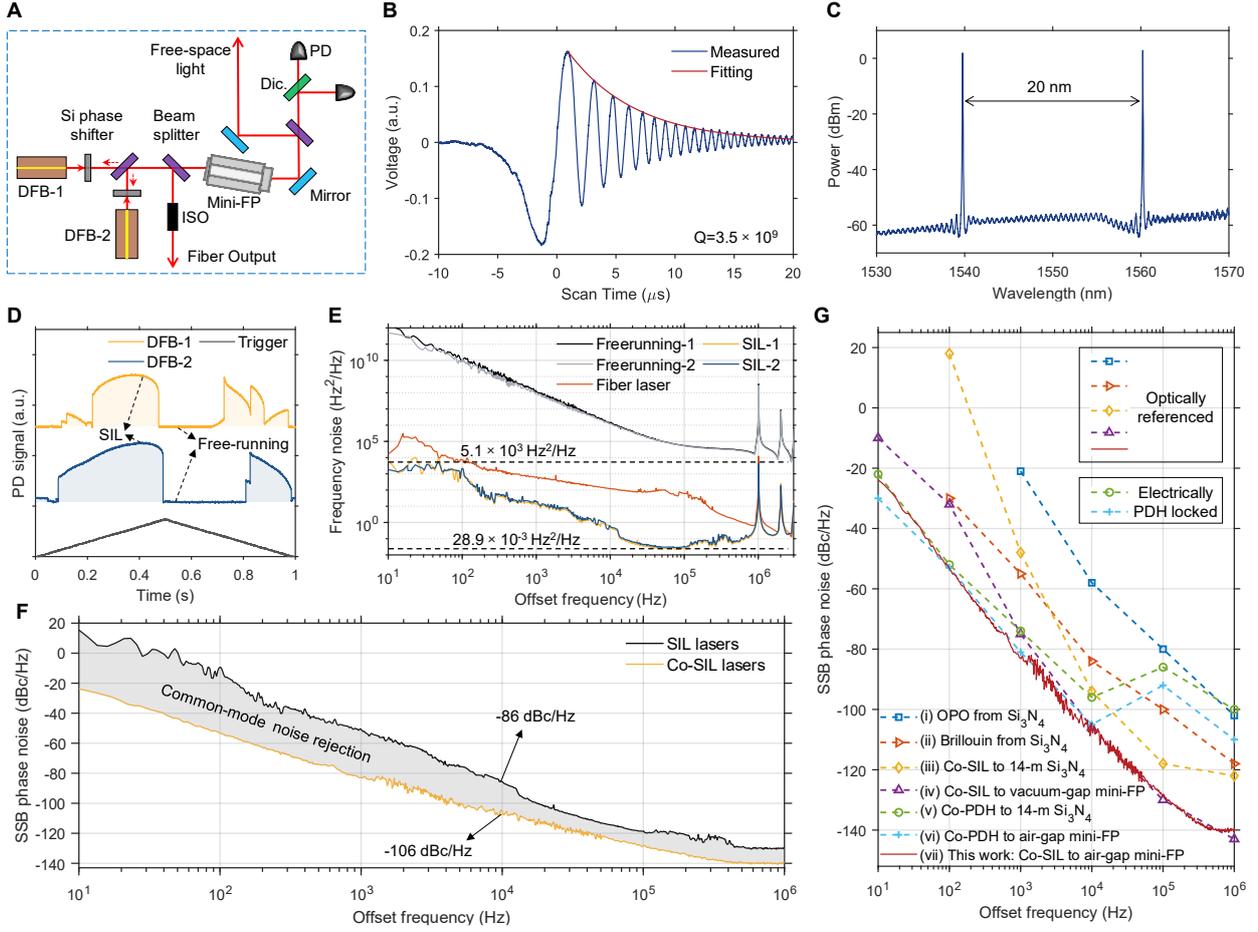

**Fig. 2. Characterization of co-SIL dual-laser references.** (**A**) Schematic of the experimental setup. The outputs of two DFB lasers are combined and split into two paths. The first path provides feedback to a mini-FP cavity for co-SIL. The transmitted light from the cavity is split: one portion is detected to monitor the locking state, while the other is output for mode profile characterization. The second path, after an isolator (ISO), provides the stabilized optical outputs. Dic., Dichroic mirror. PD, photodetector. (**B**) Ring-down trace of mini-FP reference cavity. The red curve shows an exponential fit of the oscillating amplitude. (**C**) Optical spectra of the co-SIL dual lasers. (**D**) Transmissions of the mini-FP cavity and the corresponding states of the dual DFB lasers when they are free-running and SIL. (**E**) Frequency noise power spectral density (PSD) of the dual DFB lasers under free-running and SIL states. A commercial Koheras Basik E15 fiber laser is shown for comparison. (**F**) Single-sideband (SSB) phase noise of a single SIL laser and co-SIL lasers, with the common-mode noise rejection regime between them. (**G**) Comparison of the relative phase noise performance of dual-laser references based on $Si_3N_4$ and mini-FP cavities, using either optical or electronic locking. Our co-SIL to an air-gap mini-FP cavity (curve vii) achieves lowest phase noise for an optically referenced system across a broad offset range from 100 Hz to 1 MHz. Data from: (i) OPO in a $Si_3N_4$ ring resonator (*26*), (ii) dual Brillouin lasers in a $Si_3N_4$ ring resonator (*12*), (iii) co-SIL to a 14-m $Si_3N_4$ spiral resonator (*27*), (iv) co-SIL to a vacuum-gap mini-FP cavity (*32*), and electrically PDH locked including: (v) co-PDH to a 14-m $Si_3N_4$ spiral resonator (*36*), (vi) co-PDH to an air-gap mini-FP cavity (*31*).

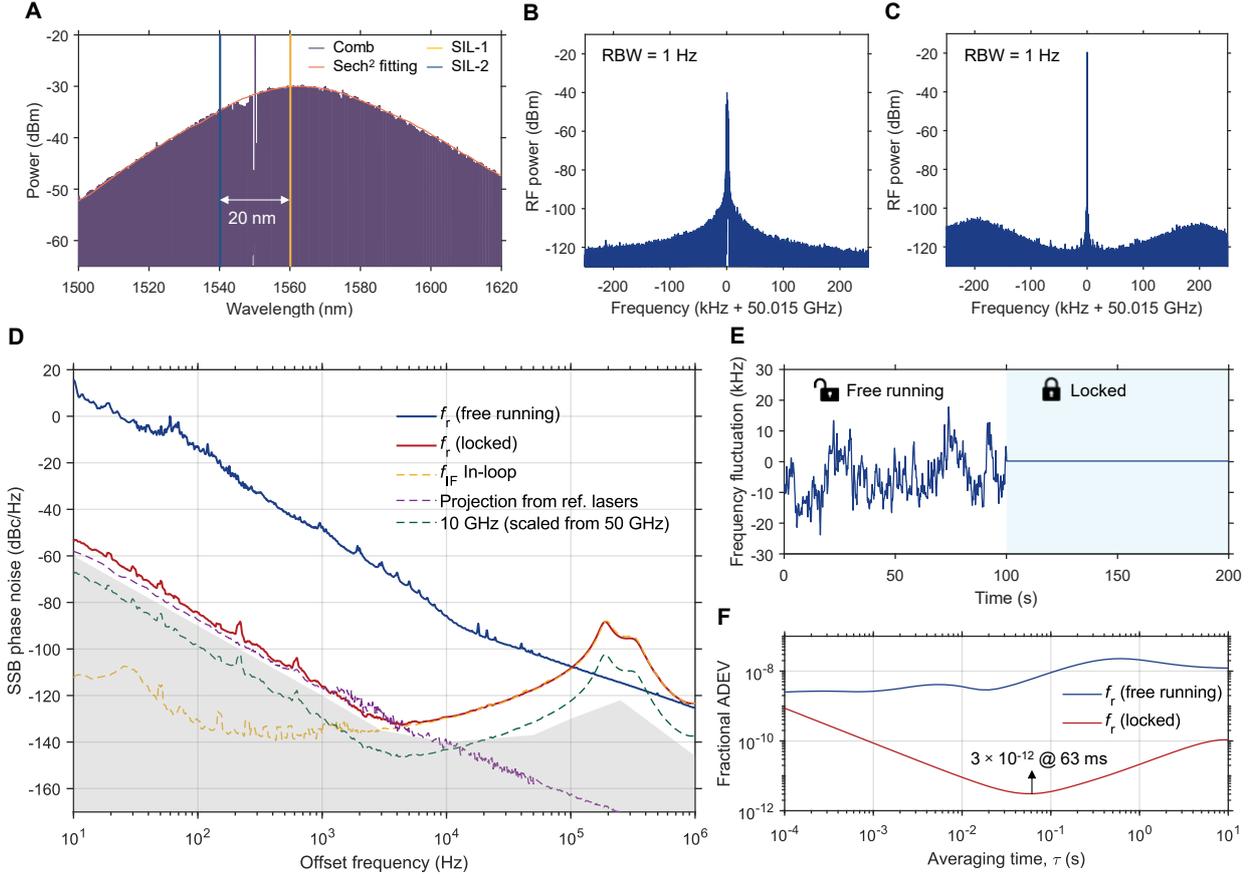

**Fig. 3. Characterization of 50 GHz (U-band) microwave generated via 2P-OFD.** (**A**) Optical spectra of the single-soliton microcomb and the co-SIL dual-laser references. The envelope of the soliton microcomb follows the sech[2] function fit overall. (**B**) Beatnote of free-running soliton microcomb, showing a broad linewidth. (**C**) Beatnote of phase-locked soliton microcomb at 50.015 GHz, showing a narrow linewidth with two servo bump shoulders near 200 kHz offset. The resolution bandwidth (RBW) of the measured microwave tone is 1 Hz. (**D**) SSB phase noise of the free-running and locked soliton microcomb beatnotes $f_r$, and the in-loop phase noise of intermediate frequency locking $f_{IF}$. The projected contribution (purple dashed curve) from the co-SIL dual-laser references, scaled down by 34 dB (i.e., $10\log_{10}(50^2)$), is shown. The phase noise of the locked $f_r$ is also scaled to a 10 GHz carrier (green dashed curve) for comparison. The grey shaded region indicates the noise floor of the phase noise analyzer. (**E**) Waterfall time trace of the repetition rate beatnote, comparing the free-running (jittery) and phase-locked (stable) 50 GHz signal. (**F**) Fractional frequency stability in terms of Allan deviation.

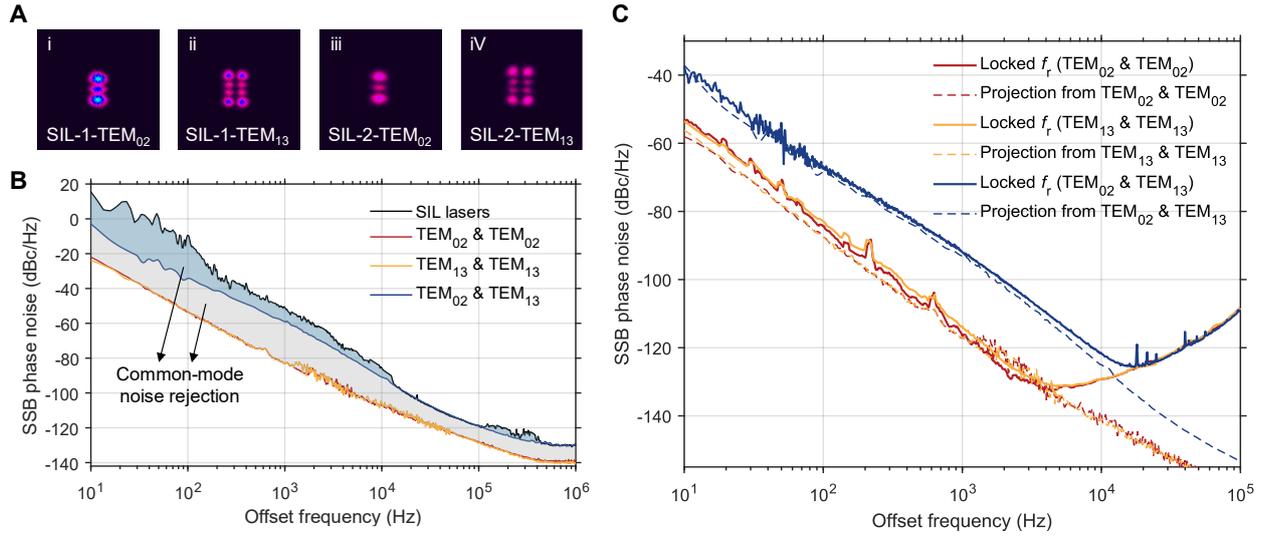

**Fig. 4. 2P-OFD performance under different co-SIL transverse mode configurations.** (**A**) Beam profile images of the free-space output from the dual-laser module when the two lasers are independently locked to two distinct high-order transverse modes of the mini-FP cavity: $TEM_{02}$ and $TEM_{13}$. (**B**) SSB phase noise of a single SIL laser and the relative phase noise between the co-SIL dual lasers under three different locking scenarios: both lasers locked to $TEM_{02}$, both to $TEM_{13}$, and one to $TEM_{02}$ and the other to $TEM_{13}$. Locking to the same transverse mode family yields identical, optimal common-mode noise rejection. (**C**) SSB phase noise of the locked 50 GHz microwave signal generated via 2P-OFD under the corresponding three co-SIL scenarios. The noise fully tracks the scaled relative phase noise of the optical references (dashed lines, scaled by 34 dB). Locking to the same mode family is crucial for achieving the lowest microwave phase noise.

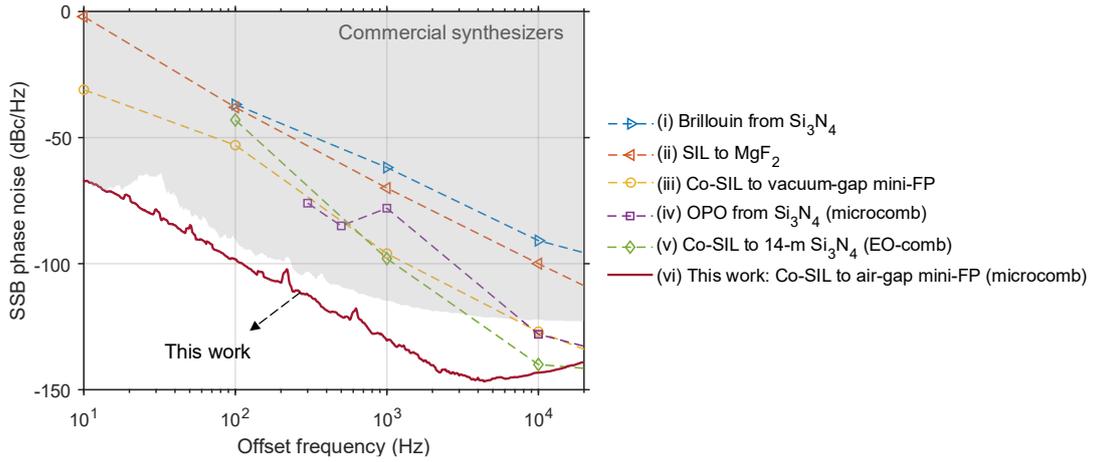

**Fig. 5. Phase noise comparison of miniature microwave oscillators based on optically locked references (all scaled to 10 GHz).** Our work (curve vi, co-SIL to an air-gap mini-FP cavity with integrated microcomb) achieves record-low phase noise among optically referenced systems at offset frequencies below 10 kHz. Data from: (i) dual Brillouin lasers from a $Si_3N_4$ ring resonator (*12*), (ii) dual lasers SIL to individual $MgF_2$ resonators (*8*), (iii) co-SIL to a vacuum-gap mini-FP cavity (*32*), (iv) OPO signal and idler from a $Si_3N_4$ ring resonator (2P-OFD with integrated microcomb) (*26*), (v) co-SIL to a 14-m spiral resonator (2P-OFD with integrated electro-optic comb (EO-comb)) (*27*).

**Table 1. Comprehensive performance comparison of miniature photonic microwave oscillators**

| Reference method | Reference cavity | Dual laser generation/ locking scheme | Frequency (GHz) | Comb type | Scaled to 10 GHz (dBc/Hz) | | | | Ref no. |
|---|---|---|---|---|---|---|---|---|---|
| | | | | | 10 Hz | 100 Hz | 1 kHz | 10 kHz | |
| Electrically PDH locked | 4-m $Si_3N_4$ | Co-PDH | 109.5 | Microcomb | -59 | -100 | -109 | -140 | (34) |
| | 14-m $Si_3N_4$ | Co-PDH | 37.3 | EO-comb | -69 | -95 | -120 | -144 | (36) |
| | Air-gap mini-FP | Co-PDH | 10 | Fiber comb | -71 | -95 | -123 | -142 | (31) |
| | Vacuum-gap mini-FP | Co-PDH | 20 | Microcomb | \ | -100 | -125 | -141 | (35) |
| | $MgF_2$ | Co-PDH | 25 | Microcomb | -93 | -115 | -131 | -149 | (25) |
| | Cylindrical mini-FP | Co-PDH | 20 | Microcomb | -75 | -101 | -133 | -152 | (57) |
| Optically referenced | $Si_3N_4$ (10.9 GHz) | Brillouin | 21.7 | \ | \ | -30 | -55 | -84 | (12) |
| | $MgF_2$ | SIL | 75-100 | \ | \ | -38 | -70 | -100 | (8) |
| | Vacuum-gap mini-FP | Co-SIL | 111.5 | \ | -31 | -53 | -96 | -127 | (32) |
| | $Si_3N_4$ (227 GHz) | OPO | 16 | Microcomb | \ | \ | -78 | -128 | (26) |
| | 14-m $Si_3N_4$ | Co-SIL | 37.7 | EO-comb | \ | -43 | -98 | -140 | (27) |
| | Air-gap mini-FP | Co-SIL | 50 | Microcomb | -67 | -99 | -131 | -144 | This work |